\documentclass[new]{JHEP3}

\usepackage{amssymb}
\usepackage{amsmath}

\usepackage{graphicx}

\renewcommand{\vec}[1]{{\bf #1}} 
\newcommand{\ii}{\'\i} 
 
\renewcommand{\vec}[1]{{\bf #1}}       
\def\beq{\begin{eqnarray}}    
\def\eeq{\end{eqnarray}}      

\def\Box{\square}                  
\def\tr{\,\mbox{tr}\,}                  
\def\Tr{\,\mbox{Tr}\,}                  
\def\Res{\,\mbox{Res}\,}                
\renewcommand{\Re}{\,\mbox{Re}\,}       
\renewcommand{\Im}{\,\mbox{Im}\,}       
\def\Lap{\Delta}                        

\def\al{\alpha}
\def\be{\beta}
\def\ch{\chi}
\def\ga{\gamma}
\def\de{\delta}
\def\ep{\epsilon}
\def\vp{\varepsilon}
\def\ze{\zeta}
\def\io{\iota}
\def\ka{\kappa}
\def\La{\Lambda}
\def\la{\lambda}
\def\na{\nabla}
\def\pa{\partial}
\def\ro{\varrho}
\def\si{\sigma}
\def\om{\omega}
\def\ph{\varphi}
\def\ta{\tau}
\def\th{\theta}
\def\te{\vartheta}
\def\up{\upsilon}
\def\Ga{\Gamma}
\def\De{\Delta}
\def\La{\Lambda}
\def\Si{\Sigma}
\def\Om{\Omega}
\def\Te{\Theta}
\def\Th{\Theta}
\def\Up{\Upsilon}

\title{Renormalization Group and Decoupling in Curved Space: \\
 II. $\,$ The Standard Model and Beyond}

\author{Eduard V. Gorbar 
\\
Departamento de F\ii sica, 
ICE, Universidade Federal de Juiz de Fora
\\ 
MG, Brazil and 
\\
Bogolyubov Institute for Theoretical Physics, Kiev, Ukraine
\\
E-mail: \email{gorbar@fisica.ufjf.br}}

\author{Ilya L. Shapiro\thanks{On leave from Tomsk State 
Pedagogical University, Tomsk, Russia.}

\\
Departamento de F\ii sica, ICE, Universidade Federal de 
Juiz de Fora
\\ 	
MG, Brazil
\\ 
E-mail: \email{shapiro@dftuz.unizar.es}}

\abstract{We continue the study of the renormalization group 
and decoupling of massive fields in curved space, started in 
\cite{apco} and analyse the higher derivative sector of the vacuum 
metric-dependent action of the Standard Model. The QCD sector at 
low-energies is described in terms of the composite effective 
fields. For fermions and scalars the massless limit shows perfect 
correspondence with the conformal anomaly, but similar limit in a 
massive vector case requires an extra compensating scalar. In all 
three cases the decoupling goes smoothly and monotonic. A 
particularly interesting case is the renormalization group flow 
in the theory with broken supersymmetry, where the sign of one 
of the beta-functions changes on the way from the UV to IR.}

\keywords{Renormalization Group, Physics of the Early Universe}
\received{March, 2003}
\preprint{DF/UFJF-03/01}

\begin{document}

$\,$

$\,$

$\,$

\section{Introduction}

\quad\quad
In the last decades there was a growing interest to 
the quantum effects of vacuum in curved space-time. In particular, 
there were investigations of the renormalization group flows for 
the parameters of the vacuum action in curved space-time: both 
perturbative and non-perturbative (see, e.g. the review \cite{flow1} 
and further references therein). At present, the most of the 
non-perturbative results concern massless (mainly supersymmetric) 
models. In some 
cases, it was possible to establish the relation between the IR 
and UV $\,\,\be$-functions for the parameters of the 
higher-derivative sector of the vacuum action.

An important application of the universal renormalization group 
flows in curved space is a candidate to be the natural model of 
inflation \cite{wave,insusy}, which is a modification of the 
well-known Starobinsky model \cite{fhh,star}. The main advantage 
of the modified Starobinsky model of inflation is that it is based 
on the first principles of quantum field theory and does not 
require conventional elements of the inflationary phenomenology 
such as inflaton, which do not have, at present, necessary particle 
physics justification. Moreover, this inflationary model 
does not require a special choice of the initial data and 
provides an automatic graceful exit to the FRW stage due to the 
supersymmetry breaking at low energies.
The theoretical background of this approach to inflation consists 
of the quantum corrections to the classical action of vacuum 
\cite{birdav,book} treated in the effective Quantum Field Theory 
framework. One of the key points is the decoupling of heavy degrees
of freedom (similar to the standard Appelquist and Carazzone 
theorem in flat space
\cite{AC}) and especially sparticles. This decoupling results in 
the transition between the stable inflationary solution at the 
beginning and the automatic graceful exit to the FRW evolution 
at the end of the inflation epoch \cite{insusy}. Qualitatively,
the decoupling is a manifestation of the dominance of the mass 
terms in the quantum effects of the massive fields at low 
energies. In the gravitational case, the masses of the fields 
compete with the energies of the gravitational quanta at the 
external lines of the vacuum diagrams (see, e.g. \cite{nova}). 
Therefore, the quantitative study of the decoupling involves the
analysis of the corresponding loop diagrams. Let us remark that,
despite the decoupling theorem is well-known for the Quantum Field 
Theory in flat space-time (see, e.g. \cite{collins,manohar}), the 
study of this phenomena in curved space-time has been started just 
recently in \cite{apco}. In this first paper we have investigated
the massive scalar field non-minimally coupled to external gravity.
The purpose of the present article is to generalize the results 
of \cite{apco} for the massive fermions and vectors, which are 
present in the low-energy spectrum in the electroweak (EW) and QCD 
sectors of the Standard Model of elementary particle physics (SM). 
Thus, we complete the description of the decoupling in the one-loop 
approximation. Furthermore, we shall discuss the decoupling of the 
sparticles in the supersymmetric theories beyond the SM. The 
present work should be considered as an important step to the 
future description of the behavior of the modified Starobinsky 
model in the interpolation regime between stable and unstable 
phases.

One of the problems in the study of the low-energy vacuum 
quantum effects of the SM is the non-perturbative nature of 
the low-energy QCD. Despite the existence of the mentioned 
above non-perturbative exact renormalization group flows in 
the massless supersymmetric version of QCD, they do not help 
too much in the realistic QCD theory which we are interested 
in here. As far as the non-perturbative results are not available 
for the phenomenologically relevant situations, we accept an 
effective field theory 
approach and describe the low-energy QCD using the Chiral 
Perturbation Theory model. It is well known, that 
all excitations of QCD at low energies are massive, and therefore 
their gravitational effects should be suppressed due to the 
decoupling mechanism. We are going to construct a qualitative 
description of this phenomena using the effective approach.

As we have seen in the previous work \cite{apco}, the calculations
on the flat background (or equivalent covariant calculations 
performed through the expansion in the powers of the
curvature tensor) can not be conclusive for the description of 
the decoupling of the cosmological constant and inverse Newton 
constant, which is unaccessible within the usual perturbative 
approach. Therefore, in this paper we will consider the 
$\be$-functions for the higher derivative terms in the effective 
action only. These terms are the most important ones for the 
inflationary model of \cite{insusy}, hence they deserve 
serious attention. 

The paper is organized as follows. In section 2 we present a 
qualitative description of the low-energy QCD and the 
corresponding effective models in curved space.
In section 3 the relation between the renormalization group for
the massive fields and conformal anomaly for the corresponding
massless fields is considered. Furthermore, we learn to make 
a distinction between various $\be$-functions in the vacuum 
sector. 
In section 4 we derive the relevant expressions for the 
effective action of massive scalars, fermions and vectors
in curved space-time. In the case of scalars and 
fermions the massless limits fit with the anomaly-induced 
effective action, while for the Proca field an extra compensating 
scalar is needed. 
In section 5 we discuss the one-loop renormalization group 
flows for these fields. The identification 
of the $\be$-functions here is slightly different from the one in
the previous paper \cite{apco}). 
In section 6 consider an application 
of these results for the model with broken supersymmetry.
Finally, in section 7 we draw our conclusions.


\section{Effective approach and the low-energy QCD in curved space}

\quad\quad
Our general purpose is to investigate the renormalization group 
flow between the UV and IR regimes in the SM in curved space-time. 
The method for 
investigating the decoupling which was developed in \cite{apco}
is essentially based on the perturbative approach and the practical 
calculations can be successfully realized only in the one loop 
approximation. In the EW sector this approximation is reliable, 
because here we meet weakly interacting fields in both UV and IR 
regimes. Indeed, many of these fields gain mass at low energies 
due to Spontaneous Symmetry Breaking and Higgs mechanism, but it 
is clear that the general approach of \cite{apco} does not meet 
big obstacles here and should be sufficient for getting the 
necessary information.

Of course, the situation in the QCD sector of the SM is 
rather different. Due to the asymptotic freedom, in the UV 
regime the perturbative description works well and hence 
there is no difficulties with the use of the method of 
\cite{apco}. The real problem is the non-perturbative nature 
of QCD at low energies because, as we have just mentioned, our 
ability to make calculations in curved space-time is mainly 
restricted by the first order of the loop expansion. Hence, 
the first task is to understand which kind of information 
about the low-energy QCD in curved space one can obtain by making 
the perturbative calculations.

As well known, the QCD Lagrangian in flat space-time reads
\begin{equation}
L\, =\, -\frac{1}{4} F_{\mu\nu}^{(a)}F^{(a)\,\,\mu\nu} 
+ \sum_{k} \left(i\bar{\psi}_k\gamma_{\mu}(\partial_{\mu} 
+ igT^aA^a_{\mu})\psi_k + m_k\bar{\psi}_k\psi_k
\right),
\label{QCD Lagrangian}
\end{equation}
where 
$\,F_{\mu\nu}^{(a)}
=\pa_{\mu}A_{\nu}^a
-\pa_{\nu}A_{\mu}^a-g f_{abc}A_{\mu }^b A_{\nu}^c\,$ 
is the field strength, $\,T^a\,$ are the $\,SU_c(3)\,$ generators, 
$\,f_{abc}\,$ are the structure constants of the $SU(3)$ algebra, 
and 
the summation is performed over the quark flavors. Due to the gauge 
symmetry the mass terms for gauge vector bosons are prohibited
such that the gluons are massless at the Lagrangian level. Switching
on the gravitational field, one may ask whether the gluons contribute 
to the running of the parameters of the gravitational action 
(\ref{vacuum 1}) at low energies, including in the modern Universe?
\footnote{We thank Natan Berkovits for this question which motivated
the discussion of this section.} In what follows we argue that the 
answer to this question is definitely no. Let us remember that, 
below the nucleosynthesis scale, the magnitude of the Hubble 
parameter is much smaller than even the masses of the lightest 
neutrinos \cite{nova}. Consequently, if we restrict our 
consideration to the semiclassical approach and therefore 
do not consider the quantum gravity effects, then only the 
photon and gluons may, in principle, contribute to the running. 

In order to study the QCD contribution to the vacuum 
gravitational renormalization group in the modern Universe, 
first of all we need to know how the QCD excitation spectrum 
gets modified when we go from 
the high energy scale down to the very low energies. 
At high energies the QCD coupling constant is small and the 
perturbation theory based on the gluons and quarks is meaningful.
Contrary to that, the low energy QCD, in terms of quarks and 
gluons (\ref{QCD Lagrangian}), is non-perturbative. At present, 
it is  not possible to determine the low-energy QCD spectrum in 
a purely theoretical framework, starting from the first principles, 
and we need a certain phenomenological input here. 
Fortunately, the QCD excitation spectrum at low energies is 
known from experiment \cite{PDG}. It is well known that there are 
no particles with the quantum numbers of gluons (needless to say 
that to explain theoretically the absence of gluons and quarks in 
the low energy QCD spectrum is a part of the confinement 
problem), and only composite particles like pions and kaons
are observed. Perhaps, it is worth reminding that due to the 
unbroken 
Poincare symmetry all states in the low-energy QCD excitation 
spectrum are characterized by their mass and spin and as usual 
one interpret these states as particles. 

And so, what we really need is the effective
spectrum of the low-energy QCD in terms of the composite 
particles. Presumably, the lightest states are the most 
important ones, because they should decouple from gravity at 
lower energies. According to the Particle Data Group \cite{PDG}, 
the lowest QCD excitations are the octet of pseudoscalar mesons:
the pions $\,\pi^0\,$ (135\, MeV) and $\,\pi^{\pm}\,$ 
(139.6 MeV), the kaons $\,K^{\pm}\,$ (493.7 MeV),
$\,\bar{K^0}\,$, and $\,K^0\,$ (497.7 MeV), and the 
$\,\eta$-meson 
$\,\eta\,$ (547.3 MeV). In massless QCD (when all bare quark 
masses are zero), these mesons would be massless since they are 
the Goldstone bosons connected with the Chiral Symmetry Breaking
(therefore, these mesons are pseudoscalars). The nine would be 
Goldstone boson $\,\eta^{\prime}\,$ acquires a nonzero mass due 
to the $\,U_A(1)\,$ axial anomaly. Since chiral symmetry is 
explicitly broken in the real QCD, the pions, kaons, and the 
$\,\eta$-meson acquire nonzero masses. For example, the 
$\,\pi^{\pm}\,$ mass is given by 
$$
m_{\pi^{\pm}}^2\,=\,-\,\frac{m_u + m_d}{f_{\pi}^2}\,<\bar{u}u>\,,
$$ 
where $\,m_u\,$ and $\,m_d\,$ are the $\,u$- and $\,d$-quark 
masses, $\,f_{\pi}\,$ is the pion decay constant, and 
$\,\,<\bar{u}u>\,\,$ is the quark condensate. In addition to 
the pseudoscalar mesons, there are also some relatively light 
scalar in the QCD spectrum like $\,f_0\,$ (600 MeV) and 
$\,a_0\,$ (980 MeV), however, these scalar states are difficult 
to resolve because of their large decay widths. The vector 
particle with the lowest mass is the $\,\rho\,$ meson. Although 
its mass $\,\rho\,$ (771.1 MeV) is larger than pseudoscalars 
masses, the $\,\rho\,$ meson plays a very important role in 
particles interactions of the low-energy QCD. Let us mention 

here only the Vector Dominance and vector meson universality 
advocated by Sakurai \cite{Sakurai}. 

The fermions with lowest 
masses in QCD are nucleons $m_p = 938.3 \,MeV$  and 
$m_n = 939.6 \,MeV$. Note that their masses are significantly 
larger than that of the pseudoscalar mesons.
Let us remark that the decoupling 
is one of the key ideas of the Heavy Baryon Chiral Perturbation 
Theory, for it allows the perturbative study of the interaction 
of nucleons and pions despite the large value of the pion-nucleon 
coupling constant. Of course, this approach is possible only at 
low energies (for details, see, e.g., \cite{Ecker}). 

According to the Chiral Perturbation Theory,
the pions and kaons are the would be Goldstone bosons
because the chiral symmetry is explicitly broken. 
Therefore, their interaction can be represented 
\cite{W,GL} as a series in $\,\,{p^2}/{16\pi^2 f_{\pi}^2}$
(where $\,p\,$ is a particle momentum), 
$\,\,{m_{\pi}^2}/{16\pi^2 f_{\pi}^2} \approx 0.01$, 
and $\,\,{m_{K}^2}/{16\pi^2 f_{\pi}^2} \approx 0.18$. 
It is convenient to collect the pseudoscalar fields in a 
unitary $\,\,3 \times 3\,\,$ 
matrix
\begin{equation}
U = \exp{(i\sqrt{2}\Phi/f_{\pi})}~,\qquad \Phi 
= \left( \begin{array}{ccc}
\dfrac{\pi^0}{\sqrt{2}} + \dfrac{\eta}{\sqrt{6}} & \pi^+ & K^+ 
\\
* \pi^-&-\dfrac{\pi^0}{\sqrt{2}}+\dfrac{\eta}{\sqrt{6}}& K^0 
\\
* K^- & \overline{K^0} & - \dfrac{2 \eta}{\sqrt{6}} 
\end{array} \right) 
\nonumber
\label{eq:phi}
\end{equation}
and the effective Lagrangian of Chiral Perturbation Theory is 
then constructed as a series in derivatives of $U$ and the 
pseudoscalars masses. For example, the leading-order term of 
the effective Lagrangian is given by
\begin{equation}
{\it L_2} = \frac{f_{\pi}}{4} 
\tr\left( \partial_{\mu}U \partial^{\mu}U^{\dagger} 
+ \chi U^{\dagger} + U\chi^{\dagger}\right),
\label{L2}
\end{equation}
where the matrix $\chi$ defines the pseudoscalars masses. 
The next-to-leading order effective Lagrangian contains
terms with four derivatives, terms quadratic in derivatives 
and linear in $\chi$, and terms quadratic in $\chi$. 

Thus, for momenta less than
$4\pi f_{\pi}$, the interaction of pions and kaons is weak and, 
consequently, it is a reasonably good approximation to consider 
one loop diagrams with free pions, kaons, and $\eta$-mesons in 
order to study the QCD contribution to the renormalization group 
running of the parameters of the vacuum gravitational action at 
low energies. Finally, since QCD does not 
have massless particles in its excitation spectrum and at energies 
less than $m_{\pi^0}$ all QCD particles decouple, 
the QCD contribution to the renormalization group running of 
the gravitational action parameters
is suppressed at low energies. In order to verify this,
in a first approximation, it is sufficient to analyse the 
contributions of free massive scalars, spinors and vectors to 
the gravitational effective action at low energies. 


\section{Renormalization group and conformal anomaly}

\quad\quad
According to the previous section, the investigation of 
decoupling of the quantized massive fields from an external 
gravity may be reduced to the derivation of the effective 
action of vacuum for those fields: scalars, fermions, and 
vecctors. In the IR, we expect that the mass terms will 
dominate and this leads to the low-energy decoupling. At high 
energies, we expect that the effects of the masses become 
negligible such that the effective action in the limit 
$\,\, m \to 0\,\,$ coincides with the effective action
derived for massless fields. In fact, this expectation 
is completely justified for the scalar and spinor fields and 
not justified at all for the massive vector. The reason is that 
the Proca field has larger number of physical degrees of freedom 
than the massless gauge vector field. This can be seen 
explicitly if one uses the Boulware transformation \cite{bou} 
for the massive vector or apply an equivalent one-loop procedure 
described in \cite{bavi}. In both cases one can see that the 
Proca field has one extra scalar degree of freedom compared
to the gauge boson.

In the MS scheme, the metric-dependent vacuum divergences 
must be removed by adding 
appropriate local counterterms and renormalizing parameters 
$\,\,\,1/G$, $\,\La$, $\,\,a_{1,2,3,4}\,\,$ of the classical 
action of vacuum 
\beq
S_{vac}\,=\,\int d^4x \sqrt{-g}\,
\Big\{\,-\frac{1}{16\pi G}\,(R+2\La)\,+\,
a_1C^2+a_2E+a_3{\Box}R+a_4R^2\,\Big\}\,,
\label{vacuum 1}
\eeq
Let us consider the massless conformal invariant fields in the
one-loop approximation. In this case only the higher derivative 
conformal part of the vacuum action (including the surface and
topological terms)
\beq
S_{HDC}\,=\,\int d^4x \sqrt{-g}\,
\left[\,a_1C^2+a_2E+a_3{\Box}R\,\right]\,,
\label{vacuum H}
\eeq
is subject to renormalization. $\,\,S_{HDC}\,$ can be called 
``conformal'', because it satisfies the Noether identity
\beq
-\, \frac{2}{\sqrt{-g}}\,
g_{\mu\nu}\,\frac{\de S_{HDC}}{\de g_{\mu\nu}}\,=\,0\,.
\label{noether1}
\eeq
On quantum level, the classical action of vacuum gains 
loop corrections and must be replaced by the corresponding 
effective action. 
Traditionally, the violation of the conformal identity
(\ref{noether1}) at quantum
level is called the conformal anomaly of the Energy-Momentum 
(stress) tensor. This anomaly has the form (see, e.g. \cite{book}
and references therein)
\beq
<T^\mu_\mu>
\,=\,\be_1^{\overline {MS}}\,\,C^2\,+\,\be_2^{\overline {MS}}\,\,E 
\,+\,\be_3^{\overline {MS}}\,\,{\Box} R\,,
\label{KG}
\eeq
where $\be_1^{\overline {MS}}$, $\be_2^{\overline {MS}}$ 
and $\be_3^{\overline {MS}}$ are the minimal subtraction
(${\overline {MS}}$-scheme) based
$\,\beta$-functions of the parameters $\,a_{1,2,3}\,$ 
correspondingly (see e.g \cite{nelspan})
 $$
 \be_1^{\overline {MS}} \,=\,-\,\frac{1}{(4 \pi)^2}\, 
 \left(\,\frac{1}{120}\,N_0 + \frac{1}{20}\,N_{1/2} 
 + \frac{1}{10}\,N_1\,\right)\,,
 $$
 $$
 \be_2^{\overline {MS}}\,=\,\frac{1}{(4 \pi)^2}\,
 \left(\,\frac{1}{360}\,N_0 + \frac{11}{360}\,N_{1/2}
 + \frac{31}{180}\,N_1\,\right)\,,
 $$
 \beq
 \be_3^{\overline {MS}}\,=\,-\,\frac{1}{(4 \pi)^2}\,
 \left(\,\frac{1}{180}\,N_0 + \frac{1}{30}\,N_{1/2} 
 - \frac{1}{10}\,N_1\,\right)\,.
 \label{oef}
 \eeq
Here $\,N_0,\,N_{1/2}\,$ and $\,N_1\,$ are the numbers
of scalar, fermion, and vector massless fields, respectively. 
Let us remark that the $\,\be$-functions 
$\,\be_2^{\overline {MS}}\,$ 
and $\,\be_3^{\overline {MS}}\,$ 
result from the renormalization of the 
topological and surface terms in (\ref{vacuum 1}). 
The corresponding divergences can be obtained 
explicitly using, for example, the Schwinger-DeWitt  
formalism (see, e.g. \cite{book}). However, in other 
calculational schemes, like the one which we shall apply
below, topological and surface divergences are not seen
explicitly. Therefore, these two $\,\be$-functions may 
be identified only through the analysis of the finite 
part of the 
effective action which is related to the anomaly (\ref{KG}). 

In the massive case, the higher-derivative divergences are 
exactly the same as for the massless fields. In the following
sections we shall apply the physical mass-dependent scheme 
of renormalization and derive corresponding physical  
$\,\,\beta$-functions which are generally different from the 
ones of (\ref{oef}). In order to distinguish the two kinds 
of the renormalization group functions, we shall denote the 
${\overline {MS}}$-scheme $\,\,\beta$-functions as 
$\,\,\beta^{\overline {MS}}$. The UV and IR limits of the 
$\,\,\beta$-functions derived in the mass-dependent scheme
will be denoted as $\,\beta^{UV}\,$ and $\,\beta^{IR}$.
Of course, we expect that the 
correctly defined $\,\be$-function would satisfy the 
relation
$$
\be^{UV}\,\,=\, \,\be(\overline{\rm MS})
\,+\, {\cal O} \Big(\frac{m^2}{p^2}\Big)\,.
$$

In this paper we shall consider $\,\be_1,$ $\,\be_3$
and $\,\be_4\,$, corresponding to the renormalization 
of the parameters $\,a_1\,$, $\,a_3\,$ and $\,a_4\,$
of the vacuum action (\ref{vacuum 1}). 
The analysis of $\,\be_2\,$ will not be presented here, 
because it requires
much more involved calculation in the third order in curvature 
\cite{bavi2}. Moreover, the decoupling in the $\,\be_2$-sector 
is less important for the application of the renormalization 
group and anomaly to inflation \cite{insusy}.

For the simple situation with the massless fields,
the expression (\ref{KG}) enables one to derive, in an
explicit and economic way, the effective action of 
gravity which is exact for the particular case of the 
homogeneous and isotropic FRW
metric. The non-local covariant form of the anomaly-induced 
effective action is \cite{rei,frts}
$$
\Gamma_{ind} \,=\, S_c[g_{\mu\nu}]\,+\, 
\frac{3\be_3^{\overline {MS}}+2\be_2^{\overline {MS}}}{36}
\,\int d^4 x \sqrt{-g (x)}\,R^2(x)\,-
$$
\beq
- \,\int d^4 x \sqrt{-g (x)}\, \int d^4 y \sqrt{-g (y)}\,
\Big[\,E - \frac23{\Box}R\,\Big]_x \,G(x,y)\,
\left[\,\frac{\be_1^{\overline {MS}}}{4}\,C^2 
+ \frac{\be_2^{\overline {MS}}}{8}
\,(E - \frac23{\Box}R)\right]_y \,.
\label{nonlocal}
\eeq
The conformal invariant functional $S_{c}[g_{\mu\nu}]$ is an 
integration constant which can not be obtained using the conformal 
anomaly. The effective
action (\ref{nonlocal}) contains a Green function $\,\,G(x,y)\,\,$
of the conformal differential fourth order scalar operator 
\beq
\De_4\,=\,{\Box}^2 + 2R^{\mu\nu}\na_\mu\na_\nu -\frac23\,R{\Box}
+\frac13\,(\na^\mu R)\na_\mu\,.
\label{pan}
\eeq

The local term $\int d^4x\sqrt{-g}R^2$ in the first line of 
(\ref{nonlocal}) gains contributions from two $\,\be$-functions 
$\,\,\be_2^{\overline {MS}}$ and $\,\,\be_3^{\overline {MS}}$. 
But, as we shall see in a moment, for our purposes only the 
$\be_3^{\overline {MS}}$-dependent contribution is relevant.  
As it was already mentioned, we are going to use the massless
effective action (\ref{nonlocal}) in order to check the more
general expression for the effective action of the massive 
fields in the UV limit. But, the calculations for the massive 
case will be performed only in the second order in curvature,
so that in order to compare the two results we have to expand 
(\ref{nonlocal}) up to the second order in curvature. 
In the situation of interest, this is equivalent to the 
bilinear expansion in the metric perturbation 
$$
h_{\mu\nu}=g_{\mu\nu}-\eta_{\mu\nu}
$$
on the flat background \cite{apco}. In both cases the operator 
(\ref{pan}) becomes just $\,\Box^2$. 
The terms with $\,C^2\,$ and $\,E\,$ in the second line of Eq. 
(\ref{nonlocal}) can be safely neglected, for they are, at least, 
of the third order. Then, the remaining term in the  second 
line of (\ref{nonlocal}) is
$$
\,\frac{\be_2^{\overline {MS}}}{18}\,
\int d^4 x \sqrt{-g (x)}\, \int d^4 y \sqrt{-g (y)}\,
({\Box}R)_x \,\Big(\frac{1}{\Box}\Big)_{(x,y)}\,({\Box}R)_y\,.
$$
After some partial integrations, this terms precisely cancels
with the $\,\be_2^{\overline {MS}}$-dependent term in the first 
line of (\ref{nonlocal}). Thus, the important consequence of our
consideration is that, in order to perform verification of the 
second-order in curvature calculations for the massive case, 
we need to take only the $\be_3^{\overline {MS}}$-dependent
part of the $\,\int d^4 x \sqrt{-g}R^2\,$-term in 
(\ref{nonlocal})
\beq
\Ga_{\be_3}\,=\,\frac{1}{12\,(4 \pi)^2}\,\int d^4 x \sqrt{-g }\,
\Big(\,-\,\frac{1}{180}\,N_0 \,-\, \frac{1}{30}\,N_{1/2} 
\,+\, \frac{1}{10}\,N_1 \,\Big) \,R^2\,,
\label{verify}
\eeq

Also, as a by-product we can see whether the quantum correction 
(\ref{nonlocal}) contributes to the propagation of the 
gravitational waves in the flat space, e.g. in the modern 
Universe. Let us notice that the $\,\int\sqrt{-g}R^2\,$ term 
does not contribute to the propagation of the transverse 
traceless (that is spin-2) mode. Therefore, the contribution 
to the equation for the  gravitational wave on the flat 
background may come only from the conformal invariant functional 
$\,S_c\,$ but not from the anomaly-induced part.

There is another aspect of the $\,\int d^4 x \sqrt{-g}R^2\,$-term,
which is important for us. The anomaly-induced effective actions 
correspond to the massless conformal fields. However, there is an 
example of the field which is massless but not conformal: the
scalar 
one with the non-minimal parameter $\,\,\xi\neq 1/6$. In this 
case the parameter $\,a_4\,$ in the vacuum action (\ref{vacuum 1}) 
has to be renormalized independently 
and there is a corresponding $\,\be$-function 
$\,\be_4^{\overline {MS}}$. On the 
other hand, the $\,\,{\Box}R$-term in the conformal anomaly 
produces the finite $\,\int \sqrt{-g}R^2$-term in the 
effective action (\ref{nonlocal}). For the massless case, these 
two $\,R^2$-terms do not mix, because the renormalization group 
is related to the divergent part of the effective action only.
For the massive fields, in a physical mass-dependent scheme,
the situation is quite different, because the $\,\be$-functions 
result
from the finite part of the effective action. Then, the division 
of the $\,\,\int\sqrt{-g} R^2$-term in the effective action between 
two $\beta$-functions: $\,\be_3\,$ for the $\,a_3\,$ parameter and 
$\,\be_4\,$ for the $\,a_4\,$ parameter in the action 
(\ref{vacuum 1}) becomes ambiguous and one has to define the 
sharing in some appropriate way. 

In the previous work \cite{apco} we attributed all the 
$\,\,\int\sqrt{-g} R^2$-term to the $\,\be_4$. In the present 
paper we  use another definition, which looks more natural. 
Since the infinite  $\,\int \sqrt{-g}R^2$-type counterterm is 
absent for $\,\xi=1/6$, we include all the terms proportional to 
$\,\,(\xi-1/6)\,\,$ into $\,\be_4$-function. All other terms 
will be attributed to the $\,\be_3$. Let us remark that this
form of the sharing is the most useful for the cosmological 
application \cite{insusy}, because the existing version of the
modified Starobinsky model is based on the supposition that  
$\,\xi \approx 1/6\,\,$. Another advantage of this definition 
is the following. We expect that in the massless limit the vacuum 
effective action of massive fields will converge to the 
anomaly-effective action. But, it is well known that the 
anomaly can be integrated only if the $\,R^2$-term in the 
anomalous trace is absent \cite{rei}. In turn, this requires 
the absence of the $\,\,\int \sqrt{-g}R^2$-counterterm, that 
can be achieved for the $\,\,\xi=1/6\,\,$ only. Hence, our 
way to identify the two $\be$-functions is natural in the 
sense it helps making the massless limit look simpler.


\section{The covariant derivation of the effective action 
up to the second order in curvature}

\quad\quad
In this section we perform the derivation of the second order 
in curvature $\,{\cal O}(R^2)$-terms in the one-loop effective 
action using the general expression for the heat kernel of the
differential second order operator derived in 
\cite{bavi2,Avramidi}. In the previous article \cite{apco} 
we have demonstrated, using the massive scalar field as an example,
that this covariant heat-kernel approach is completely equivalent 
to the calculation of the quantum correction to the graviton 
propagator from the matter loop (see also \cite{basti}
for the similar calculation). 
Hence, our approach is analogous to the standard study of 
decoupling in QED \cite{manohar}. 

We define the one-loop Euclidean effective action of a field 
with mass $\,m$  as a trace of an integral 
of the heat kernel over the proper time $\,s$ (compared to 
\cite{bavi2} there is an important $\exp[-m^2s]$ insertion)
\beq
{\bar \Ga}^{(1)}\,=\,-\,\frac12\,{\Tr}{\ln}
\Big(\,-{\na}^2{\hat 1}+m^2-{\hat P}+\frac16\,R{\hat 1}\Big)
\,\,=\,\,\,-\,\frac12\,\int_{0}^{\infty}\,
\frac{ds}{s}\,\tr\,K(s)\,,
\label{effective}
\eeq
where $\,K(s)\,$ is the heat kernel
$$
\tr\,K(s)\,=\,\frac{(\mu^2)^{2-w}}{(4\pi s)^w}\,
\int d^4x\,g^{1/2}\,e^{-sm^2}\,{\rm tr}\,\left\{\,{\hat 1}
\,+\,s{\hat P}
+\,s^2\,\Big[ R_{\mu\nu}f_1(-s\na^2)R^{\mu\nu}
\,+
\right.
$$
\beq
\left.
+ Rf_2(-s\na^2)R + {\hat P}f_3(-s\na^2)R
+ {\hat P}f_4(-s\na^2){\hat P}
+ {\hat {\cal R}}_{\mu\nu}f_5(-s\na^2){\hat {\cal R}}^{\mu\nu}
\,\Big]
\right\}\,+\,{\cal O}({\cal R}^3)\,,
\label{heat}
\eeq
where $\,{\hat {\cal R}}_{\mu\nu}=\Big[\na_\mu\,,\,\na_\nu\Big]\,$
is a commutator of covariant derivatives in the space of the 
fields of interest. The functions $f_{1,2,..,5}(\tau)$ are given
by the following expressions: 
$$
f_1(\tau)=\frac{f(\tau)-1+\tau/6}{\tau^2}
\,,\,\,\,\,\,\,\,\,\,\,\,\,\,\,\,\,\,\,
f_2(\tau)=\frac{f(\tau)}{288}+\frac{f(\tau)-1}{24\tau}
-\frac{f(\tau)-1+\tau/6}{8\tau^2}\,,
$$
$$
f_3(\tau)=\frac{f(\tau)}{12}
+\frac{f(\tau)-1}{2\tau}
\,,\,\,\,\,\,\,\,\,\,\,\,\,\,\,\,\,\,\,
f_4(\tau)=\frac{f(\tau)}{2}
\,,\,\,\,\,\,\,\,\,\,\,\,\,\,\,\,\,\,\,
f_5(\tau)=\frac{1-f(\tau)}{2\tau}\,,
$$
where 
$$
f(\tau)=\int_0^1 d\al\,e^{\al(1-\al)\tau}\,,\,\,\,\,\,\,\,\,\,\,
\,\,\,\,\,\,\,\,\,\,\, \tau=-s\na^2\,.
$$
Below we consider how these formulas can be applied to the 
derivation of the effective action of scalars, fermions (in this
case the overall sign in the Eq. (\ref{effective}) has to be 
changed, of course), and vectors.

\subsection{Massive scalar with the non-minimal coupling} 

Let us first repeat the calculation for the scalar field
performed in the previous paper \cite{apco}. The main 
difference is that here we perform more detailed analysis 
of the massless UV limit for the non-conformal case 
$\,\,\xi \neq 1/6$.

For the massive real scalar with the non-minimal coupling 
to gravity 
\beq
{\hat P}=-(\xi-1/6)R
\,\,\,\,\,\,\,\,\,\,\,\,\,\,\,\,\,\,\,\,
{\rm and}
\,\,\,\,\,\,\,\,\,\,\,\,\,\,\,\,\,\,\,\,
{\hat {\cal R}}_{\mu\nu}=0\,.
\label{P-scalar}
\eeq

Introducing the new variable $sm^2=t$ and notation
$\,u={\tau}/{t}$, we arrive at the 
following integral representation for the effective action:
$$
{\bar \Ga}^{(1)}\,=\,-\,\frac{1}{2(4\pi)^2}\,\int d^4x g^{1/2}\,
\left(\frac{m^2}{4\pi \mu^2}\right)^{w-2}\int_{0}^\infty
dt\,e^{-t}\,\left\{\,
\frac{m^4}{t^{w+1}}\,+\,\Big(\xi-\frac16\Big)\,\frac{R\,m^2}{t^w}
\,+
\right.
$$
\beq
\left.
+\,\sum_{i=1}^{5}\,l^*_i\cdot R_{\mu\nu} M_i R^{\mu\nu}
\,+\,\sum_{j=1}^{5}\,l_j\cdot R M_j R\,\right\}\,,
\label{intermediate}
\eeq
where 
$$
l_{1,2}^*=0   \,,\,\,\,\,\,\,\,\,\,\,\,\,\,\,\,\,
l_3^*=1       \,,\,\,\,\,\,\,\,\,\,\,\,\,\,\,\,\,
l_4^*=\frac16 \,,\,\,\,\,\,\,\,\,\,\,\,\,\,\,\,\,
l_5^*=-1 \,;
$$
\vskip 1mm
$$
l_1=\frac{1}{288}-\frac{1}{12}\,\Big(\xi-\frac16\Big)
+\frac{1}{2}\,\Big(\xi-\frac16\Big)^2
\,,\,\,\,\,\,\,\,\,\,\,\,
l_2=\frac{1}{24}-\frac{1}{2}\,\Big(\xi-\frac16\Big)\,,
$$
\vskip 1mm
$$
l_3=-\frac{1}{8}
\,,\,\,\,\,\,\,\,\,\,\,\,
l_4=-\frac{1}{16}+\frac{1}{2}\,\Big(\xi-\frac16\Big)
\,,\,\,\,\,\,\,\,\,\,\,\,
l_5=\frac{1}{8}
$$
and 
\beq
M_1=\frac{f(tu)}{u^2t^{w+1}}\,,\,\,\,\,\,\,\,\,\,
M_2=\frac{f(tu)}{ut^w}\,\,,\,\,\,\,\,\,\,\,\,
M_3=\frac{f(tu)}{u^2t^{w+1}}\,,\,\,\,\,\,\,\,\,\,
M_4=\frac{1}{ut^w}\,,\,\,\,\,\,\,\,\,\,
M_5=\frac{1}{u^2t^{w+1}}\,.
\label{M}
\eeq

As we shall see later on, the representation (\ref{intermediate})
exists also for the fermion and vector fields. The difference 
is just the values of the coefficients $l_i^*$ and $l_j$. Hence,
from practical point of view it is better to derive the 
integrals $\,\int_{0}^\infty dt\,e^{-t}\,M_i(t,u)\,$ only 
once and later use them as a standard reference. 

Following Barvinsky and Vilkovisky \cite{bavi2}, we adopt the 
dimensional regularization in the form suggested by Brown and 
Cassidy \cite{brocas} (see also \cite{bavi} for useful technical 
details). The UV limit $\tau/sm^2 \gg 1$ and the IR limit 
$\tau/sm^2 \ll 1$ can be easily investigated in this framework.
The relevant integrals are
$$
\,\Big(\frac{m^2}{\mu^2}\Big)^{\om-2}\,
\int_{0}^\infty \frac{dt}{(4\pi)^\om}\,M_1(t)
\,=\,
\,\Big(\frac{m^2}{4\pi \mu^2}\Big)^{w-2}\,
\int_{0}^\infty \frac{dt}{(4\pi)^2}\,e^{-t}\,t^{1-\om}\,
\int_0^1 d\al\,e^{\al(1-\al)\,tu}\,=
$$
\beq
=\,\frac{1}{(4\pi)^\om}\,\Big[\,\frac{1}{2-\om}\,+\,
\ln\,\Big(\frac{m^2}{4\pi \mu^2}\Big)\,+\,2A\,\Big] \,,
\label{M1}
\eeq
\vskip 1mm
$$
\,\Big(\frac{m^2}{\mu^2}\Big)^{\om-2}\,
\int_{0}^\infty \frac{dt}{(4\pi)^\om}\,M_2(t)\,=\,
\,\Big(\frac{m^2}{4\pi \mu^2}\Big)^{w-2}\,
\int_{0}^\infty \frac{dt}{(4\pi)^2}\,e^{-t}\,t^{-\om}\,
\frac{1}{u}\,
\int_0^1 d\al\,e^{\al(1-\al)\,tu}\,=
$$
\beq
=\,\frac{1}{(4\pi)^\om}\,\Big\{\,\Big[\frac{1}{2-\om}\,+\,
\ln\,\Big(\frac{m^2}{4\pi \mu^2}\Big)+1\Big]\cdot
\Big(\frac{1}{12}-\frac{1}{a^2}\Big)\,-\,\frac{4A}{3a^2}
\,+\,\frac{1}{18}\,\Big] \,,
\label{M2}
\eeq
\vskip 1mm
$$
\,\Big(\frac{m^2}{\mu^2}\Big)^{\om-2}\,
\int_{0}^\infty \frac{dt}{(4\pi)^\om}\,M_3(t)\,=\,
\,\Big(\frac{m^2}{4\pi \mu^2}\Big)^{w-2}\,
\int_{0}^\infty \frac{dt}{(4\pi)^2}\,e^{-t}\,t^{-1-\om}
\,\frac{1}{u^2}\,\int_0^1 d\al\,e^{\al(1-\al)\,tu}\,=
$$
\beq
=\,\frac{1}{(4\pi)^\om}\,\Big\{\,\Big[\,\frac{1}{2-\om}\,+\,
\ln\,\Big(\frac{m^2}{4\pi \mu^2}\Big)+\frac32\,\Big]\cdot
\Big(\frac{1}{2a^4}-\frac{1}{12a^2}+\frac{1}{160}\Big)
\,+\,\frac{8A}{15a^4}\,-\,\frac{7}{180a^2}\,+\,\frac{1}{400}
\,\Big]\,,
\label{M3}
\eeq
\vskip 1mm
$$
\,\Big(\frac{m^2}{\mu^2}\Big)^{\om-2}\,
\int_{0}^\infty \frac{dt}{(4\pi)^\om}\,M_4(t)\,=\,
\,\Big(\frac{m^2}{4\pi \mu^2}\Big)^{w-2}\,
\int_{0}^\infty \frac{dt}{(4\pi)^2}\,e^{-t}\,t^{-\om}\,
\frac{1}{u}\,=
$$
\beq
=\,\frac{1}{(4\pi)^\om}\,\cdot\,\Big[\,\frac{1}{2-\om}
\,+\,
\ln\,\Big(\frac{m^2}{4\pi \mu^2}\Big)+1\,\Big]\,\cdot
\,\Big(\,\frac{1}{4}-\frac{1}{a^2}\,\Big)\,,
\label{M4}
\eeq
and
$$
\,\Big(\frac{m^2}{\mu^2}\Big)^{\om-2}\,
\int_{0}^\infty \frac{dt}{(4\pi)^\om}\,M_5(t)\,=\,
\,\Big(\frac{m^2}{4\pi \mu^2}\Big)^{w-2}\,
\int_{0}^\infty \frac{dt}{(4\pi)^2}
\,e^{-t}\,t^{-1-\om}\,\frac{1}{u^2}\,=
$$
\beq
=\,\frac{1}{(4\pi)^\om}\,\cdot\,\Big[\,\frac{1}{2(2-\om)}
\,+\,\frac12\,\ln\,\Big(\frac{m^2}{4\pi \mu^2}\Big)
\,+\,\frac34\,\Big]\,\cdot
\,\Big(\,\frac{1}{4}-\frac{1}{a^2}\,\Big)^2\,,
\label{M5}
\eeq
where
\beq
A =-\frac12\int_0^1 d\al\ln\Big[1+\al(1-\al)u\Big]
=1-\frac{1}{a}\ln \frac{1+a/2}{1-a/2}
\,\,\,\,\,\,\,\,\,\,\,\,\, {\rm and} \,\,\,\,\,\,\,\,\,\,\,\,\,
a^2=\frac{4\na^2}{\na^2-4m^2}\,.
\label{A}
\eeq
In the expressions (\ref{M1}) - (\ref{M5}) we disregarded those 
terms which vanish in the $\om \to 2$ limit.
Now one has to replace these integrals into Eq. (\ref{intermediate})
and use the relation \cite{bavi2}\footnote{This relation holds
only in the second order in curvature, and also means that we 
disregard the topological Gauss-Bonnet term. In other words, it 
can be used if we are interested in the corrections to the 
gravitational propagator, but not to the vertex terms.}
\beq
R_{\mu\nu}\,{\widehat X}\,R^{\mu\nu}
\,-\,\frac13\,R\,{\widehat X}\,R \,=
\,\frac12\,C_{\mu\nu\al\be}\,{\widehat X}\,C^{\mu\nu\al\be} 
\,+\,{\cal O}(R^3)\,,
\label{C}
\eeq
valid for any operator $\,{\widehat X}\,$ 
build up from the powers of the covariant
derivative. In this way,  we arrive at the effective 
action for massive nonminimal scalar in the ${\cal O}(R^2)$
approximation \cite{apco}
$$
{\bar \Ga}^{(1)}_{scalar}
\,=\,\frac{1}{2(4\pi)^2}\,\int d^4x \,g^{1/2}\,
\left\{\,\frac{m^4}{2}\cdot\Big[\frac{1}{2-w}
+\ln \Big(\frac{4\pi \mu^2}{m^2}\Big)+\frac32\Big]\,+
\right.
$$$$
\left.
+\,\Big(\xi-\frac16\Big)\,m^2R\,
\Big[\,\frac{1}{2-w}
+\ln \Big(\frac{4\pi \mu^2}{m^2}\Big)+1\,\Big]\,+
\right.
$$$$
\left.
+\,\frac12\,C_{\mu\nu\al\be} \,\Big[\,\frac{1}{60\,(2-w)}
\,+\,\frac{1}{60}\ln \Big(\frac{4\pi \mu^2}{m^2}\Big)+k_W(a)
\,\Big] C^{\mu\nu\al\be}\,+
\right.
$$
\beq
\left.
+\,R \,\Big[\,\frac12\,\Big(\xi-\frac16\Big)^2\,
\Big(\,\frac{1}{2-w}
+ \ln \Big[\frac{4\pi \mu^2}{m^2}\Big]\,\Big)
+ k_R(a)\,\Big]\,R\,\right\}\,,
\label{final}
\eeq
where 
$$
k_W(a)\, = \,\frac{8A}{15\,a^4}
\,+\,\frac{2}{45\,a^2}\,+\,\frac{1}{150}\,,
$$
\beq
k_R(a)\, =
\,A\Big(\xi-\frac16\Big)^2-\frac{A}{6}\,\Big(\xi-\frac16\Big)
+\frac{2A}{3a^2}\,\Big(\xi-\frac16\Big)
+\frac{A}{9a^4}-\frac{A}{18a^2}+\frac{A}{144}+
\nonumber
\\
+\frac{1}{108\,a^2}
-\frac{7}{2160} + \frac{1}{18}\,\Big(\xi-\frac16\Big)\,.
\label{W}
\eeq

An important check of the expression (\ref{final}) can be 
performed through the massless limit $m\to 0$, where we expect 
to meet the ${\cal O}(R^2)$ part of the anomaly-induced 
effective action (\ref{nonlocal}). It is easy to see that 
the limit $\,m\to 0\,$ is singular for the expression $A$ in 
(\ref{A}). Fortunately, for $\,\xi=1/6$, the $\,A$-dependent 
terms which contain this singularity cancel and we obtain 
\beq
{\bar \Ga}^{(1)}(\xi=1/6,\,m\to 0)
\,=\,-\,\frac{1}{12\cdot 180(4\pi)^2}\,
\int d^4x \,g^{1/2}\,R^2 \,+...\,.
\label{anomaly-induced}
\eeq
This is exactly what we should expect due to (\ref{verify}). 
Hence, the free scalar field with $\,\xi=1/6\,$ has consistent 
UV limit. But, let us remind that the nonminimal parameter 
$\,\xi\,$ can not be equal to $\,1/6\,$ precisely within the
interacting theory, because at higher loops such theory would  
be non-renormalizable \cite{hath}. On the other hand, if we
set $\,\xi\neq 1/6\,$ and perform the massless limit, the 
overall $\,(\xi-1/6)^2\int \sqrt{-g} R^2\,$ term in the 
effective action has regular behavior due to the cancelation
with the $\,\,\,\ln \Big[{4\pi \mu^2}/{m^2}\Big]\,\,\,$ term.
The singularity really appears not in the $\,m^2\to 0\,$
limit, but in the high-energy $|p^2|\to \infty$ limit. 
Of course, the coefficient of this divergence is identical
to the pole $\,\,1/(\om-2)$ coefficient. Hence,
the singularity of the expression $\,A\,$ in the 
$\,m^2\to 0\,$ or $|p^2|\to \infty$ limits is related to
the UV divergence and must be treated by renormalization.
Technically, this singularity shows the relation between
the divergent and finite parts of the effective action in 
the UV. Similarly, in the massless limit $\,a\to 2\,$ 
there are 
terms proportional to $\,A\,$ in the Weyl term formfactor 
$\,k_W(a)$. The UV singularities in $\,k_W(a)$ and 
$\,k_R(a)\,$ for $\,\xi \neq 1/6\,$ mean the appearance 
of the covariant non-local terms like 
\beq
\int d^4x\,\sqrt{-g}\,C_{\mu\nu\al\be}\,
\ln \Big(-{\Box}/{\mu^2}\Big)
C_{\mu\nu\al\be} \,\,\,\,\,\,\,\,\,\,\,\,\,\,\,\,\,
{\rm and} \,\,\,\,\,\,\,\,\,\,\,\,\,\,\,\,\,
\int d^4x \,\sqrt{-g}\,R\,
\ln \Big(-\Box/{\mu^2}\Big)R \,.
\label{EH}
\eeq
These terms correspond to the renormalization group running, 
which we are going to discuss in section 5.

\subsection{Massive fermion theory}

In this case the differential operator of interest is 
\beq
{\hat H}_f\,=\,i \Big(\gamma^\mu\nabla_\mu\,+\,im_f\Big)\,,
\label{differential}
\eeq
where $\,m_f\,$ is a fermion mass.
After the standard doubling procedure (see, e.g. 
\cite{book}
and also \cite{peixo} for a proof of the equality
$\,\,
\Tr\ln\,[\gamma^\mu\nabla_\mu\,+\,im_f]
\,=\,\Tr\ln\,[\gamma^\mu\nabla_\mu\,-\,im_f]\,$)
$\,$ 
and using the relations
$$
{\hat {\cal R}}_{\mu\nu}\psi
\,=\,\Big[\na_\mu\,,\,\na_\nu\Big]\psi
\,=\,-\,\frac14\,R_{\mu\nu\al\be}
\ga^\al\ga^\be\,\psi
\,\,\,\,\,\,\,\,\,\,\,\,\,\,\,\,\,\,\,\,\,\,
{\rm and}
\,\,\,\,\,\,\,\,\,\,\,\,\,\,\,\,\,\,\,\,\,\,
\ga^\mu\ga^\nu\na_\mu\na_\mu\,=\,\na^2\,-\,\frac14\,R\,,
$$
we arrive at the following coefficients in the massive 
fermion case:
\beq
l_1 = 0\,,\,\,\,\,\,
l_2 = -\frac{1}{16}\,,\,\,\,\,\,
l_3 = -\frac18\,,\,\,\,\,\,
l_4 = \frac{1}{24}\,,\,\,\,\,\,
l_5 = \frac{1}{8}\,.
\label{l1-5 fermion}
\eeq
\beq
l_1^* = 0\,,\,\,\,\,\,
l_2^* = \frac14\,,\,\,\,\,\,
l_3^* = 1\,,\,\,\,\,\,
l_4^* = -\frac{1}{12}\,,\,\,\,\,\,
l_5^* = -1\,.
\label{l1-5ast fermion}
\eeq
Of course, the sign of the whole expression (\ref{heat}) must 
be changed due to the fermion statistics of the field. After 
we established the coefficients (\ref{l1-5 fermion}) and 
(\ref{l1-5ast fermion}), the calculation reduces to the 
routine application of the equations (\ref{M1}) - (\ref{M5}).
The effective action has the form 
$$
{\bar \Ga}^{(1)}_{fermion}
\,=\,\frac{1}{2(4\pi)^2}\,\int d^4x \,g^{1/2}\,
\left\{\,-2\,m_f^4\cdot\Big[\frac{1}{2-w}
+\ln \Big(\frac{4\pi \mu^2}{m_f^2}\Big)+\frac32\Big]\,+
\right.
$$$$
\left.
+\,\frac{1}{3}\,m_f^2R\cdot\Big[\,\frac{1}{2-w}
+\ln \Big(\frac{4\pi \mu^2}{m_f^2}\Big)+1\,\Big]\,+
\right.
$$
\beq
\left.
+\,\frac12\,C_{\mu\nu\al\be} \,\Big[\,\frac{1}{10\,(2-w)}
\,+\,\frac{1}{10}\ln \Big(\frac{4\pi \mu^2}{m_f^2}\Big)+k^f_W(a)
\,\Big] C^{\mu\nu\al\be}\,\,\,+\,\,\,R \,\Big[\, k^f_R(a)\,\Big]
\,R\,\right\}\,,
\label{final-f}
\eeq
where 
$$
k^f_W(a)\, = \,\frac{300Aa^2-480\,A 
\,-\, 40 a^2 \,+\, 19\,a^4}{225\,a^4}\,,
$$
\beq
k^f_R(a)\, =
\frac{a^4 - 120A - 10a^2 + 30Aa^2}{270\,a^4}\,.
\label{Cf}
\eeq
Of course, in the expressions for $\,\,a\,\,$ and $\,\,A\,\,$ from 
(\ref{A}), one has to replace the scalar mass $\,\,m\,\,$ by the 
fermion one $\,\,m_f$. Let us remark that the $\,R^2$-type
divergence in (\ref{final-f}) is absent, because the divergent 
term does not depend on the presence of a mass and because the
massless theory is conformally invariant. 

Finally, in the massless limit $\,\,m_f\to 0\,\,$ 
we obtain 
\beq
{\bar \Ga}^{(1)}(m_f\to 0)
\,=\,-\,\frac{1}{12\cdot 30(4\pi)^2}\,
\int d^4x \,g^{1/2}\,R^2 \,+...\,,
\label{anomaly-induced fermion}
\eeq
in a perfect correspondence with the expected result (\ref{verify}).

\subsection{Massive vector theory}

The massive vector case (Proca theory) is the most complicated
one, this especially concerns the correspondence with the 
anomaly-induced result (\ref{verify}) in the massless limit.
Let us remark that this correspondence represents a very 
complete test of the calculations. The resulting 
$\,\int\sqrt{-g}R^2$-term is sensible to the contributions from 
absolutely all higher derivative terms in the effective action
for the massive theory. Those terms which do not contribute to 
(\ref{verify}) directly are verified through the cancelation 
of the singular $A$-structures or through the correspondence 
with the pole terms. 

The massive vector operator is 
\beq
{\hat H}_v\,=\,\de_\nu^\mu \Box\, - \,\na^\mu\na_\nu
\,-\, R^\mu_\nu \,-\, \de_\nu^\mu \,m^2_v\,,
\label{vector o}
\eeq
and one can show that \cite{bavi}
\beq
\Tr \ln {\hat H}_v\,=\,
\Tr \ln \Big(\de_\nu^\mu \Box - R^\mu_\nu 
\,-\, \de_\nu^\mu\,m^2_v\Big)
\,-\, \Tr \ln \Big(\Box - m^2_v\Big)
\,+\, \ln m^2\,\int d^4x\,\de^\mu_\mu\,\de(0)\,,
\label{vector oo}
\eeq
where the first term is the non-degenerate massive vector 
operator and the second term is a scalar operator 
which is required to compensate an extra degree of freedom 
of the non-degenerate massive vector compared to the Eq. 
(\ref{vector o}). Since the third term in (ref{vector oo})
does not depend on curvature, we will omit it in what follows.

As we have just mentioned, the efficient check of the calculations
can be done by taking the massless limit and consequent comparison 
with the Eq. (\ref{verify}). However, for the vector case the 
massless limit can not be simple\footnote{For instance, this
can be seen from the third term in (\ref{vector oo}).}. 
In fact, we have to link 
the massless limit of the expression (\ref{vector oo}) with the 
similar, but different formula for the massless vector:
\beq
\Tr \ln {\hat H}_v(m_v\equiv 0)\,\,=\,\,
\Tr \ln \Big(\,\de_\nu^\mu \,\Box - R^\mu_\nu\,\Big)
\,\,-\,\, 2\Tr \ln \,\Box\,.
\label{gauge vector}
\eeq
In the massless case, the original theory is gauge invariant 
and the formula (\ref{gauge vector}) results from the 
DeWitt-Faddeev-Popov procedure. The second term in the 
{\it r.h.s.} of the last equation represent the contribution
of the Faddeev-Popov gauge ghosts. Of course, 
there are two scalar (ghost) degrees of greedom in the gauge 
field case\footnote{As far as we are interested in the vacuum
gravitational effect, there is no difference between Abelian 
and non-Abelian theories. For the sake of simplicity, we shall
consider the simplest Abelian version only.}, in contrast to the 
single compensating scalar in the massive vector theory.
Hence, the receipt of how to check the expression for the 
effective gravitational action of a massive vector field is the 
following.
First, one has to derive the contribution of both non-degenerate 
vector and minimal scalar fields and then use (\ref{vector o}).
But, in order to have a correspondence with the conformal anomaly 
in the massless limit, the scalar contribution
must be multiplied by the factor of two.

Now we are in a position to perform the calculations. In the
vector case
$$
{\hat {\cal R}}_{\mu\nu}
\,=\,[\,{\hat {\cal R}}_{\mu\nu}\,]^\al_\be
\,=\,-\,R^\al_{\,\,\,\be\mu\nu}
\,,\,\,\,\,\,\,\,\,\,\,\,\,\,\,\,\,\,\,\,\,
{\hat P}\,=\,P^\nu_\mu\,=\,\frac16\,R\,\de^\nu_\mu - R^\nu_\mu\,.
$$
After a simple algebra we obtain the following coefficients: 
\beq
l_1 = -\frac18\,,\,\,\,\,\,
l_2 = -\frac12\,,\,\,\,\,\,
l_3 = -\frac12\,,\,\,\,\,\,
l_4 = \frac{5}{12}\,,\,\,\,\,\,
l_5 = \frac12\,,\,\,\,\,\,
\label{l1-5}
\eeq
\beq
l_1^* = \frac12\,,\,\,\,\,\,
l_2^* = 2\,,\,\,\,\,\,
l_3^* = 4\,,\,\,\,\,\,
l_4^* = -\frac43\,,\,\,\,\,\,
l_5^* = -4\,.
\label{l1-5ast}
\eeq
Using (\ref{l1-5}),
(\ref{l1-5ast}) and the Eq. (\ref{intermediate}), 
we arrive at the effective action of a massive vector field
$$
{\bar \Ga}^{(1)}_{vector}
\,=\,\frac{1}{2(4\pi)^2}\,\int d^4x \,g^{1/2}\,
\left\{\,\frac{3}{2}\,m_v^4\cdot\Big[\frac{1}{2-w}
+\ln \Big(\frac{4\pi \mu^2}{m_v^2}\Big)+\frac32\Big]\,+
\right.
$$$$
\left.
+\,\frac12\,m_v^2R\,\Big[\,\frac{1}{2-w}
+\ln \Big(\frac{4\pi \mu^2}{m_v^2}\Big)+1\,\Big]\,+
\right.
$$
$$
\left.
+\,\frac12\,C_{\mu\nu\al\be} \,\Big[\,\frac{13}{60\,(2-w)}
\,+\,\frac{13}{60}\ln \Big(\frac{4\pi \mu^2}{m_v^2}\Big)+k^v_W(a)
\,\Big] C^{\mu\nu\al\be}\,+
\right.
$$
\beq
\left.
+\,R \,\Big[\,
\frac{1}{72\,(2-w)}
\,+\,\frac{1}{72}\,\ln \Big(\frac{4\pi \mu^2}{m_v^2}\Big)
\,\,+\,\, k^v_R(a)\,\Big]
\,R\,\right\}\,,
\label{final-v}
\eeq
where
$$
k^v_W(a)\, = 
\,-\frac{91}{450}+\frac{2}{15a^2}
-\frac{8A}{3a^2}+A+\frac{8A}{5a^4}\,,
$$
\beq
k^v_R(a)\,=\,-\,\frac{1}{2160}+\frac{A}{48}
+\frac{A}{3a^4}+\frac{1}{36a^2}-\frac{A}{18a^2}\,.
\label{Cv}
\eeq
Here, $a$ and $A$ in Eq. (\ref{A}) depend on the vector mass 
$m_v$. The divergent part of (\ref{final-v}) coincides with the 
expressions derived in \cite{bavi}.
It is easy to see that the massless limit of the formfactor 
$\,\,k^v_R(a)\,\,$ is singular, because the $\,A$-dependent terms
do not cancel. This is exactly what we should expect, because the
consistent massless limit requires subtracting one more scalar 
contribution.
In case of the double compensating scalar, instead of (\ref{Cv}),
we meet the following formfactors in the higher derivative sector
(here, index $gv$ means ``gauge-like vector'')
$$
k^{gv}_W(a)\, = 
\,\frac{241}{3600}-\frac{5A}{16}
-\frac{A}{5a^4}-\frac{1}{60a^2}+\frac{5A}{6a^2}\,,
$$
\beq
k^{gv}_R(a)\, =
\frac{13}{1080} - \frac{A}{24} + \frac{1}{54a^2}    
+\frac{2A}{9a^4}+\frac{A}{9a^2}\,.
\label{Cg}
\eeq
Using these formfactors, in the massless limit $m_v\to 0$ we 
meet a non-singular result
and obtain 
\beq
{\bar \Ga}^{(1)}(m_v\to 0)
\,=\,+\,\frac{1}{12\cdot 10(4\pi)^2}\,
\int d^4x \,g^{1/2}\,R^2 \,+...\,,
\label{anomaly-induced vector}
\eeq
in a perfect correspondence with (\ref{verify}).

It is clear that the last result can not be interpreted such 
that we achieved a non-singular massless limit of the Proca 
field contribution to the effective action. But, in this way 
we checked both non-degenerate vector and compensating scalar 
contributions and thus ensured the correctness of our result 
(\ref{final-v}), (\ref{Cv}). 


\section{Renormalization group equations}

\quad\quad
The purpose of this section is to derive the renormalization 
group $\be$-functions for the parameters $\,a_1$, $\,a_3$,
and $\,a_4\,$ of the vacuum action (\ref{vacuum 1}) in the 
mass-dependent scheme. This calculation is relevant for the 
anomaly-induced inflation model of \cite{insusy}, because 
this model is based on the phenomena of the decoupling of 
the massive fields at low energies. Thus, we can 
not be completely satisfied by the standard mass-independent 
$\be$-functions which arise in the Minimal Subtraction
($\overline{\rm MS}$) scheme. In fact, the 
$\be^{(\overline{\rm MS})}$-functions 
describe the running only at high energies, when they 
correspond to the leading logarithms in the vacuum 
effective action. The corresponding expressions 
\cite{buvo,Maroto} are the gravitational analogs of 
the generalized Euler-Heisenberg effective action in
QED. An important advantage of the $\overline{\rm MS}$
scheme is that the renormalization group can be formulated 
in a completely consistent nonperturbative covariant way 
\cite{tmf,nelspan,Toms,buvo,book}, 

while the renormalization of the vacuum action in the 
mass-dependent scheme is not so general and has to involve
the covariant expansion in the curvature tensor
or (as we did in the previous paper \cite{apco}) 
an expansion of the metric around the fixed (flat) 
background. However, in
the high-energy UV limit the $\overline{\rm MS}$ scheme 
and the mass-dependent scheme $\be$-functions must coincide
and this enables one to check the envolved calculations in 
the mass-dependent scheme.

In the $\overline{\rm MS}$ scheme the $\be$-function
of the effective charge $C$ is defined as 
\beq
\be_C(\overline{\rm MS})
\,=\,\lim_{n\to 4}\,\mu\,\frac{dC}{d\mu}\,.
\label{beta}
\eeq
The derivation of the $\be$-functions for the parameters 
$a_1$, $a_3$, and $a_4$ of the vacuum action (\ref{vacuum 1}),
in the mass-dependent scheme, has been described in 
\cite{apco} (see, e.g. \cite{manohar} for the general 
technical introduction to the effective approach in 
Quantum Field Theory). On flat background, one has to 
subtract the 
counterterm at the momentum $p^2=M^2$, where $M$ is the 
renormalization point and calculate the $\be$-function 
using the formula
\beq
\be_C = \lim_{n\to 4}\,M\,\frac{dC}{dM}\,.
\label{beta-mass-M}
\eeq
Instead, one can simply take the derivative $\,-pd/dp\,$ 
of the formfactors in the polarization operator. In the 
covariant formalism, we identify $\,p^2\,$ and 
$\,-\Box\,$ and rewrite
the definition above using the variable $\,\,a\,\,$ 
from the Eq. (\ref{A}), therefore, the $\,\be$-functions
are the operators in the $x$-space. Now we can apply 
this procedure 
to the formfactors of the $\,C_{\mu\nu\al\be}^2\,$ and 
$\,R^2\,$ terms in the scalar, fermion, and vector cases.


\subsection{Massive scalar}

\quad
\quad
In this case the $\,\be_1$-function has the form \cite{apco}
\beq
\be_1^{scalar} \,=\,-\, \frac{1}{(4\pi)^2}\, \Big(\, 
\frac{1}{18a^2}\,-\,\frac{1}{180}\,-\, 
\frac{a^2-4}{6a^4}\,A\,\Big)\,,
\label{beta-mass}
\eeq
that is the general result for the one-loop $\be$-function
valid at any scale. As one should expect, the $\,\be$-function 
for the Weyl term coefficient $\,a_1\,$ does not depend on the 
nonminimal parameter $\,\xi$.

As the special cases we meet the UV limit $p^2 \gg m^2$
\beq
\be_1^{scalar,\, UV}\,=\,-\,\frac{1}{(4\pi)^2}\,\frac{1}{120} 
+ {\cal O}\Big(\frac{m^2}{p^2}\Big)\,,
\label{beta-UV}
\eeq
that agrees with the $\overline{\rm MS}$-scheme result 
(\ref{oef}). In the IR limit 
$\,p^2 \ll m^2\,$ we meet 
\beq
\be_1^{scalar,\, IR}\,=\,-\,\frac{1}{1680\,(4\pi)^2}\,\cdot\,
\frac{p^2}{m^2} \,\,
+ \,\,{\cal O}\Big(\frac{p^4}{m^4}\Big)\,.
\label{beta-IR}
\eeq
The last formula demonstrates the IR decoupling of the 
quantum effects of the massive scalar field. Moreover, 
the decoupling is a smooth monotone effect, that can be 
seen from the first plot (i) at the Figure 1. In this plot
as in all similar plots in other figures, we use the 
variable $\,a\,$ defined in (\ref{A}). The advantage of 
the use of $\,a\,$ is that this variable changes from 
$\,a=0\,$  in the
IR to $a=2$ in the UV, while $p$ changes from $p=0$ to
$p=\infty$. It is important that the dependence 
$\,a^2={4p^2}/(p^2+4m^2)\,$ on $\,p^2\,$ is also 
monotonic.
\begin{center}
\begin{figure}[tb]
\begin{tabular}{cc}
\mbox{\hspace{1.5cm}} (i) & \mbox{\hspace{0.5cm}} (ii) \\
\mbox{\hspace{1.0cm}}
\resizebox{!}{4cm}{\includegraphics{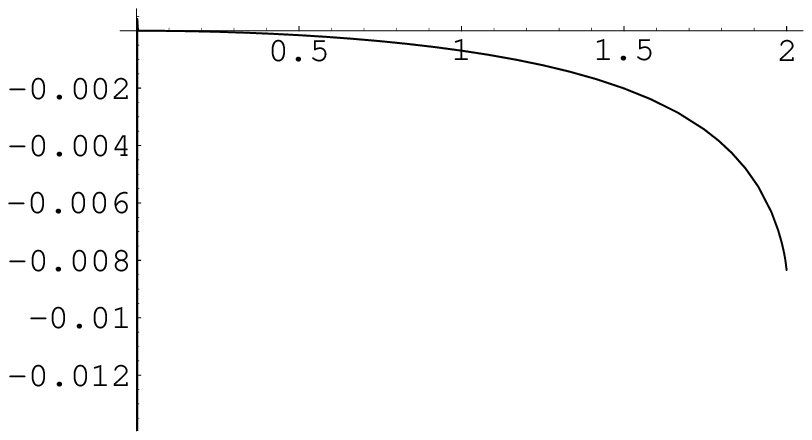}}
& \resizebox{!}{4cm}{\includegraphics{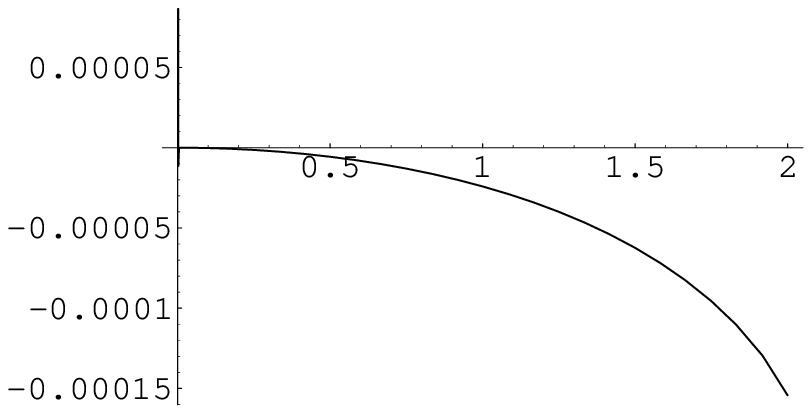}}
\end{tabular}

\vskip 3mm

\begin{quotation}

\noindent \textbf{Figure 1}.
\textsl{ The plots of the $\beta$-functions
$\,$ (i) $\,$ $\be^{scalar}_1$ $\,$ and 
$\,$ (ii) $\,$ $\be_3^{scalar}$ versus $\,a$.
All the $\,\be$-functions in this and consequent plots
are presented in the units of $\,(4\pi)^2$.}

\end{quotation}

\label{fig1}

\end{figure}

\end{center}


Let us consider the remaining $\,\,\be$-functions $\,\be_3\,$ 
and $\,\be_4$. According to the definition given in section 3, 
$\,\be_4\,$ is defined through the same procedure as $\,\be_1$,
if we separate the $\,(\xi-1/6)\,$-proportional terms in the 
formfactor $\,\,k_R(a)\,$ in Eq. (\ref{W}). Direct calculations 
give the following expression (see also \cite{apco})
\beq
\be_4^{scalar}
\,=\,-\,\frac{1}{(4\pi)^2}\,\Big(\xi-\frac16\Big)\,
\Big\{\,\frac{1}{8}\,(\,4A\,-\,a^2A\,+\,a^2\,)
\,\Big(\xi-\frac16\Big)\,+\,\frac{a^2-4}{48}\,\cdot\,
\Big(\,\frac{a^2-12}{a^2}\,A\,\,-1\,\Big)\,\Big\}\,.
\label{beta-2 mass scalar}
\eeq
In the UV limit $\,\,p^2 \gg m^2\,\,$ the $\,\be$-function 
is (in agreement with the standard $\,\,{\overline{MS}}\,$ 
result \cite{nelspan})
\beq
\be_4^{scalar,\,UV}
\,\,=\,\,-\,\frac{1}{2(4\pi)^2}\,\Big(\xi-\frac16\Big)^2 
+ {\cal O}\Big(\frac{m^2}{p^2}\Big)\,,
\label{beta 2-UV scalar}
\eeq
while in the IR limit $p^2 \ll m^2$ we obtain
\beq
\be_4^{scalar,\,IR} \,\,=\,\,-\, \frac{1}{12\,(4\pi)^2} 
\,\,\Big[\,\Big(\xi-\frac16\Big)^2
-\frac{1}{15}\,\Big(\xi-\frac16\Big)
\,\Big]\,\cdot\,\frac{p^2}{m^2}
\,\,+\,\, {\cal O}\Big(\frac{p^4}{m^4}\Big)\,.
\label{beta 2-IR scalar}
\eeq
In order to define the $\,\be_3$-function, according to the 
definition given in the section 3, one do not need to take 
derivative $\,\,M\,d/dM$. Instead, we can use the definition 
$$
-\,\frac{2}{\sqrt{-g}}\,g_{\mu\nu}\,
\frac{\de\,\Ga}{\de \,g_{\mu\nu}}\,=\,<T^\mu_\mu>\,,
$$
and the relation between the quantum-corrected trace and
the $\,\beta$-functions, which we generalize from the massless
case\footnote{It is, of course, simpler to {\it define} the 
$\,\be_3$-function as a $\,(\xi-1/6)\,$-independent part 
of the $R^2$-sector of the effective action and do not  
mention the trace anomaly which plays only an illustrative 
role here. Our way of presentation is motivated by the 
will to maintain the link to the anomaly-induced 
effective action.}. 
Thus, the required $\,\be_3$-function can be defined 
directly from the $\,(\xi-1/6)\,$-independent part of the 
formfactor $\,\,k_R(a)\,$ in Eq. (\ref{W}):
\beq
\be_3^{scalar}\,=\,
\frac{1}{(4\pi)^2}\,\Big[\,\frac{20-7\,a^2}{360\,a^2}
\,+\,\frac{(a^2-4)^2\,A}{24\,a^4}\,\Big]\,.
\label{beta 3}
\eeq
The UV limit of this renormalization group function 
perfectly corresponds to the standard ${\overline {MS}}$
result (\ref{oef})
\beq
\be_3^{scalar, \,UV}\,=\,-\,
\frac{1}{180\,(4\pi)^2}\,+{\cal O}\Big(\frac{m^2}{p^2}\Big)\,,
\label{beta 3 UV}
\eeq
while the IR limit demonstrates the decoupling of the scalar
field
\beq
\be_3^{scalar, \,IR}\,\,=\,\,-\,\,
\frac{1}{1260\,(4\pi)^2}\,\,\frac{p^2}{m^2}
\,\,+\,\,{\cal O}\Big(\frac{p^4}{m^4}\Big)\,.
\label{beta 3 IR}
\eeq
Finally (this is very important for the inflationary model
of \cite{wave}), the decoupling goes smoothly and is
monotonic, as one can see from the second plot at the
Figure 1.


\subsection{Massive fermion}

\quad\quad
The $\be$-function of the $a_1$ coefficient has the form
\beq
\be_1^{fermion}\,=\, \frac{1}{(4\pi)^2}\, \Big[\, 
\frac{2}{9a^2}\,-\,\frac{19}{180}\,+\,
\Big(\,\frac{8}{3a^4}\,-\,\frac{5}{3a^2}
\,+\,\frac{1}{4}\,\Big)\,A\,\Big]\,.
\label{beta-mass-f}
\eeq
The UV limit $p^2 \gg m^2$ gives
\beq
\be_1^{fermion,\,UV}\,=\,-\,\frac{1}{20\,(4\pi)^2}\,
+ \,{\cal O}\Big(\frac{m^2}{p^2}\Big)
\label{beta-UV-f}
\eeq
in agreement with the $\overline{\rm MS}$-scheme result
and also with the expression for divergences in the effective 
action (\ref{final-f}). The IR limit $p^2 \ll m^2$
is qualitatively similar to the scalar case in the 
sense that it shows the decoupling 
\beq
\be_1^{fermion,\,IR}\,=\,-\,\frac{1}{168\,(4\pi)^2}
\,\cdot\,\frac{p^2}{m^2} \,\,
+ \,\,{\cal O}\Big(\frac{p^4}{m^4}\Big)\,.
\label{beta-IR-f}
\eeq
The plot demonstrating the smooth and monotone form 
of the decoupling for the $\,\be_1\,$ is the first one 
(i) at the Figure 2. 


\begin{center}

\begin{figure}[tb]

\begin{tabular}{cc}

\mbox{\hspace{1.5cm}} (i) & \mbox{\hspace{0.5cm}} (ii) \\

\mbox{\hspace{1.0cm}}
\resizebox{!}{4cm}{\includegraphics{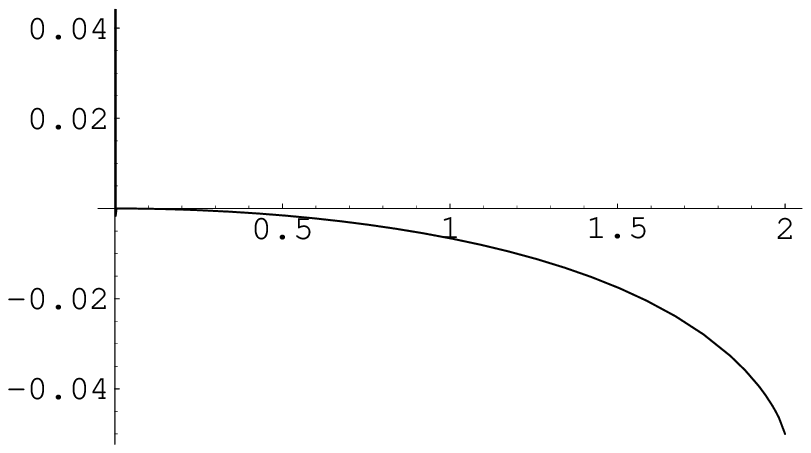}}

& \resizebox{!}{4cm}{\includegraphics{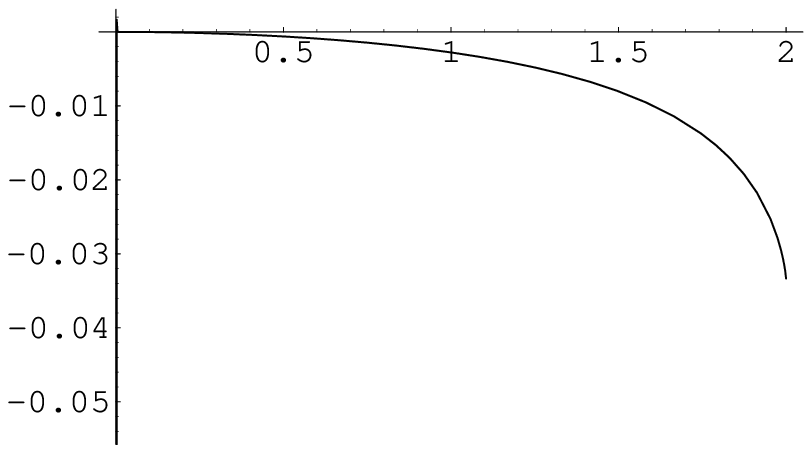}}

\end{tabular}

\vskip 3mm

\begin{quotation}

\noindent \textbf{Figure 2}.
\textsl{ The plots of the $\beta$-functions
(i) $\be^{fermion}_1$  and 
 (ii) $\be_3^{fermion}$ versus $a$.}
\end{quotation}

\label{fig2}

\end{figure}

\end{center}


According to our definition in section 3, the $\,\be_4$-function 
is identically zero for the fermion field. The derivation of the 
$\,\be_3$-function performs similarly to the scalar 
case, but for the fermions we have to use the formfactor 
$\,\,k^f_R(a)\,\,$ from (\ref{Cf}). The result is 
\beq
\be_3^{fermion}\,=\,
\frac{1}{(4\pi)^2}\,\Big[\,\frac{a^2-10}{45\,a^2}
\,+\,\frac{2\,(a^2-4)\,A}{3\,a^4}\,\Big]\,.
\label{beta 3 fermion}
\eeq
The UV limit of this renormalization group function 
corresponds to the expected standard ${\overline {MS}}$
result (\ref{oef})
\beq
\be_3^{fermion, \,UV}\,=\,-\,
\frac{1}{30\,(4\pi)^2}\,+\,{\cal O}\Big(\frac{m^2}{p^2}\Big)\,,
\label{beta 3 UV fermion}
\eeq
while the IR limit  demonstrates the decoupling of the 
spinor field
\beq
\be_3^{fermion, \,IR}\,\,=\,\,-\,\,
\frac{1}{420\,(4\pi)^2}\,\cdot\,\frac{p^2}{m^2}
\,\,+\,\,{\cal O}\Big(\frac{p^4}{m^4}\Big)\,.
\label{beta 3 IR fermion}
\eeq
The second plot at the Figure 2 shows that the decoupling 
goes smoothly and is monotonic.

\vskip 2mm
\subsection{Massive vector}

\quad\quad
Despite the massless limit for the quantum vacuum 
corrections of the massive vector is singular
(due to the gauge invariance and consequently different 
number of physical degrees of freedom), we can
derive the general expressions for the $\,\be$-functions
using the usual receipt. Then, the $\be$-function for the 
$\,a_1\,$ coefficient is
\beq
\be_1^{vector}\,=\, \frac{1}{(4\pi)^2}\, \Big[\, 
\frac{11}{60}\,-\,\frac{1}{6a^2}\,-\,\frac{a^2}{16}
\,+\,\frac{(a^2-4)(a^4-8a^2+8)}{16\,a^4}\,\cdot\,A\,\Big]\,.
\label{beta-mass-v}
\eeq

The UV limit $p^2 \gg m^2$ demonstrates perfect 
correspondence
with the divergent coefficient of the Weyl term in the
effective action (\ref{final-v})
\beq
\be_1^{vector,\,UV}\,=\,-\,\frac{1}{(4\pi)^2}\,\frac{13}{120} 
+ {\cal O}\Big(\frac{m^2}{p^2}\Big)\,.
\label{beta-UV-v}
\eeq
Of course, this corresponds, also, to the standard 
$\overline{\rm MS}$-scheme result \cite{bavi}.
Compared to the UV limit, exactly as in the scalar and 
fermion cases, the IR regime $p^2 \ll m^2$  demonstrates  
the decoupling of the loop contribution
\beq
\be_1^{vector,\,IR}\,=\,-\,\frac{3}{\,112\,(4\pi)^2}
\,\cdot\,\frac{p^2}{m^2} \,\,
+ \,\,{\cal O}\Big(\frac{p^4}{m^4}\Big)\,.
\label{beta-IR-v}
\eeq
The decoupling occurs in a smooth manner, according to the 
first plot at the Figure 3. 

\begin{center}
\begin{figure}[tb]
\begin{tabular}{cc}
\mbox{\hspace{1.5cm}} (i) & \mbox{\hspace{0.5cm}} (ii) \\
\mbox{\hspace{1.0cm}}
\resizebox{!}{4cm}{\includegraphics{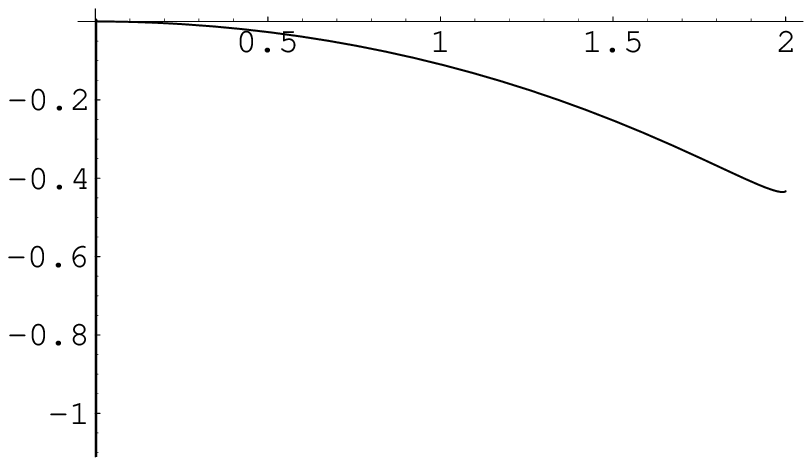}}
& \resizebox{!}{4cm}{\includegraphics{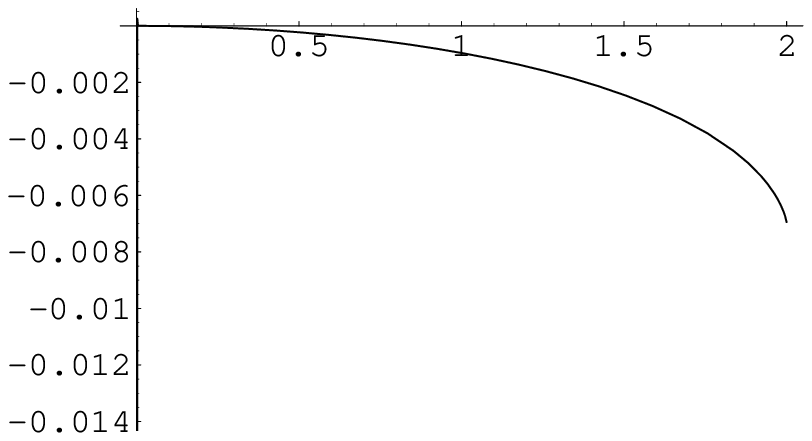}}
\end{tabular}

\vskip 3mm

\begin{quotation}

\noindent \textbf{Figure 3}.
\textsl{The plots of the $\beta$-functions
$\,$ (i) $\,$ $\be_1(a)$ $\,$ and $\,$ (ii) $\,$ $\be_4(a)$ 
for the case of the massive vector field.}
\end{quotation}

\label{fig3}

\end{figure}

\end{center}


In the $\,R^2\,$- sector we meet a usual problem with the 
separation of the $\,\be_3\,$ and $\,\be_4\,$ functions.
The problem is that, we can not use the same definition 
as for the scalar case, because in the vector case
there is no parameter 
similar to $\,\xi$. At the same time, the presence of the 
$\,\int\sqrt{-g}R^2\,$ divergence and the consequent UV
singularity in the finite $\,R^2\,$- term (of course, this
only means that the non-local renormalization-group 
related term (\ref{EH}) emerges) does not permit us 
to attribute all the $\,R^2\,$- term in the effective action
to $\,\be_3\,$. Hence, the explicit criterion for the 
separation is really absent. But, this does not indicate 
an inconsistency of the theory. Let us remind
that the object of principal physical importance is 
the effective action (\ref{final-v}), which has no 
ambiguity beyond the usual renormalization point dependence.
The separation of the two $\,\be$-functions means that we 
separate, in a certain manner, the finite and infinite 
$\,R^2\,$-terms. Then, fixing the ambiguity in the 
$\,\be$-functions we define the certain framework for the 
well defined object such as effective action.

After we apply the standard procedure, the expression 
for the $\,\be_4\,$-function has the form
\beq
\be_4^{vector}
\,=\,\frac{1}{768\,a^2\,(4\pi)^2}\,
\Big[\,\frac{( a^2-4)\,(80 - 8\,a^2 + a^4)}{a^2} \,
\cdot\, A
\,\,-\,\,\frac{80 - 16\,a^2 + 3\,a^4 }{3}
\,\Big]\,.
\label{beta 3 vector}
\eeq
In the UV limit we meet 
\beq
\be_4^{vector,\,UV}
\,=\,-\,\frac{1}{144\,(4\pi)^2}\,,
\label{beta 3 vector UV}
\eeq
in accordance with the corresponding divergence of 
the effective action (\ref{final-v}), hence we agree with the
${\overline {MS}}$-scheme in this limit. In the IR limit the
result is 
\beq
\be_4^{vector,\,IR}
\,\,=\,\,-\, \frac{1}{1120\,(4\pi)^2} \,\cdot\,\frac{p^2}{m^2}
\,\,+\,\, {\cal O}\Big(\frac{p^4}{m^4}\Big)\,,
\label{beta 3 vector IR}
\eeq
that demonstrates the usual decoupling. The plot of the
 $\,\be_3\,$-function is presented at Figure 3.

\section{Decoupling in the case of supersymmetry}

\quad\quad
According to the analysis of the previous section, the 
$\,\be$-functions show the standard
IR behavior, that is the smooth and monotonic decoupling.
It is interesting to see whether one can meet a similar feature 
of the $\,\be$-functions in case of decoupling
of the supersymmetric partners of the observable sector of 
the supersymmetric Standard Model (MSSM) or other supersymmetric 
gauge theory such as GUT. In the one-loop approximation, the 
$\,\be$-functions gain independent additive 
contributions from all the fields. We know that, for the 
energy scale below the mass of the particles, these 
$\,\be$-functions start to decrease and for 
$\,p^2 \ll m^2$ the $\,\be$-functions are quadratically 
suppressed.  Therefore, 
in order to investigate the details of the superpartners 
decoupling, we need to know the masses of all these 
superpartners. Unfortunately, this information is not 
available, and hence there is no chance to perform the 
detailed quantitative analysis of the decoupling. At the 
same time, it is possible to construct a very simple
model which can illustrate the principal characteristics 
of supersymmetry decoupling.

Let us consider, as an example, the minimal supersymmetric 
extension (MSSM) of the Standard Model (SM).
The particle content of the SM includes 12 vectors 
(gluons, photon, $W$ and $Z$ vector bosons), quarks and 
leptons, which can be counted, in terms of the Dirac 
spinors, as $\,N_{quark}=18$ and $\,N_{lepton}=6\,$ 
(we suppose that the neutrino are Dirac massive particles, 
in case they are Maiorana particles this number gets changed 
$\,N_{lepton}=4.5$, but all the conclusions remain the same),
a complex Higgs doublet, which is equivalent to four real 
scalars. In total, using the notations of (\ref{oef}), we have 
\beq
N_0^{SM}=4\,,\,\,\,\,\,\,\,\,\,
N_{1/2}^{SM}=24\,,\,\,\,\,\,\,\,\,\,
N_{1}^{SM}=12\,.
\label{SM}
\eeq
With this particle content we meet a positive overall sign 
of the induced $\,\int\sqrt{-g}R^2$-term in (\ref{verify}).
The same sign takes place in the present-day Universe, where 
only photon is active. And this sign has a very strong 
physical meaning. If we do not introduce large negative 
$a_4$ coefficient in the classical action (\ref{vacuum 1}), 
the positive sign in (\ref{verify}) means non-stability of
the anomaly-induced inflation \cite{star}. On the 
contrary, if this sign is negative, the inflation is stable.
Indeed, this happens in the MSSM, where
\beq
N_0^{MSSM}=104\,,\,\,\,\,\,\,\,\,\,
N_{1/2}^{MSSM}=32\,,\,\,\,\,\,\,\,\,\,
N_{1}^{MSSM}=12\,.
\label{MSSM}
\eeq
As it was suggested in the first reference of \cite{insusy}, 
the decoupling of the 
sparticles may be responsible for the change of sign in 
(\ref{verify}), and this provides an appealing scheme of 
the modified Starobinsky model of inflation which starts 
in the stable regime due to supersymmetry and has a 
natural graceful exit to the FRW evolution after 
supersymmetry breaks down.  

Our purpose is to obtain the qualitative description of 
the decoupling, without involving the details of the 
supersymmetric spectrum. Therefore, let us assume the 
simplest possible input. Suppose that, for some 
reason, {\it all} the
sparticles have masses much larger than the masses of the 
observed particles\footnote{In some sense, this is a 
natural hypothesis. If there is no unknown general law of 
nature of this kind, it is very difficult to explain why 
the observed particles can not be superpartners of each 
other. Also, it is worth mentioning that this spectrum 
of supersymmetry leads to a natural inflation!}. For the 
sake of simplicity, we suppose that all
the constituents of the SM as massless. Moreover, we
shall simplify things further and suppose that all the
sparticles have exactly the same mass, which we denote 
$\,M^*$. Hence, we arrive at the ``super-simplified'' model 
with the number of massless fields given by (\ref{SM}) 
and with the number of fields with an equal mass $\,M^*$
given by the difference with (\ref{MSSM}). An additional 
advantage 
of our simplifications is that one does not have 
massive vectors in this model and hence we are free 
from the corresponding problems with their massless limit 
which have been discussed in sections 4 and 5. 

At energies higher than the EW scale,
the massless SM fields contribute as
$$
\be_1^{SM}\,=\,N_{0}^{SM}\,\cdot\, 
\be_1^{scalar,\overline {MS}}
\,\,+\,\,N_{1/2}^{SM}\,\cdot\,\be_1^{fermion,\overline {MS}}
\,\,+\,\, N_{1}^{SM}\,\cdot\,\be_1^{vector,\overline {MS}}
$$
and their massive superpartners as 
$$
\be_1^{SUSY}\,=\,
(N_0^{MSSM}-N_0^{SM})\,\cdot\,\be_1^{scalar}(a)
\,\,+\,\,(N_{1/2}^{MSSM}-N_{1/2}^{SM})\,\cdot\,\be_1^{fermion}(a)
\,+
$$$$
+\,(N_{1}^{MSSM}-N_{1}^{SM})\,\cdot\,\be_1^{vector}(a)\,,
$$
where the parameter $\,a\,$ depends on the mass
$\,M^*$.
The overall $\,\be$-function for the parameter $a_1$ is 
given by the sum 
\beq
\be_1^t\,=\,\be_1^{SM}\,+\,\be_1^{SUSY}\,=\,
\frac{1}{(4\pi)^2}
\,\left[\,\frac{2\,A\,(a^2-4)\,(3a^2+17)}{3\,a^4}
\,-\,\frac{49a^2+68}{18\,a^2}\,\right]\,.
\label{SUSY-Weyl}
\eeq
The UV and IR limits of this expression are given by 
\beq
\be_1^{t\,UV}\,=\,-\,\frac{1}{(4\pi)^2}\,\cdot \frac{11}{3}
\,+\,{\cal O}\Big(\frac{m^2}{p^2}\Big)\,,
\,\,\,\,\,\,\,\,\,\,\,\,\,\,\,\,\,\,
\be_1^{t\,IR}\,=\,-\,\frac{1}{(4\pi)^2}\,\cdot\,
\left(\,\frac{73}{30}
\,+\,\frac{3}{28}\,\frac{p^2}{m^2}\,\right)
\,+\,{\cal O}\left(\frac{p^4}{m^4}\right)\,,
\label{SUSY-Weyl IR and UV}
\eeq
respectively, and the plot of the $\,\be$-function is 
presented at the Figure 4. As usual, the decoupling is 
smooth and monotone.

\begin{center}

\begin{figure}[tb]

\begin{tabular}{cc}

\mbox{\hspace{1.5cm}} (i) & \mbox{\hspace{0.5cm}} (ii) \\

\mbox{\hspace{1.0cm}}
\resizebox{!}{4cm}{\includegraphics{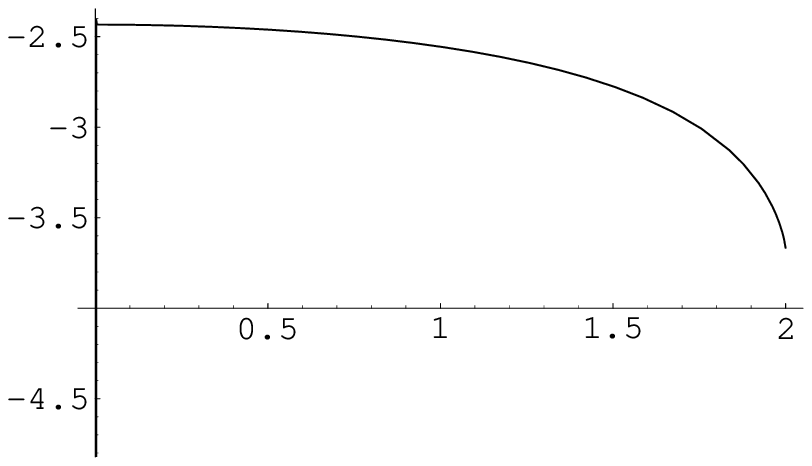}}

& \resizebox{!}{4cm}{\includegraphics{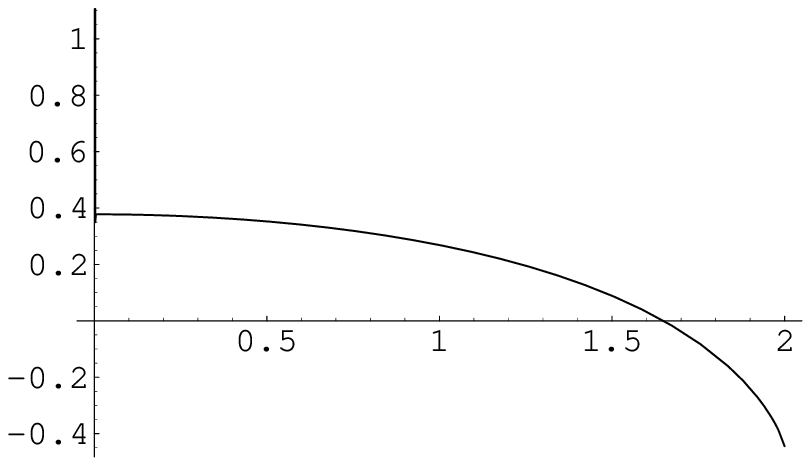}}

\end{tabular}

\vskip 3mm

\begin{quotation}

\noindent \textbf{Figure 4}.
\textsl{The plots of the $\beta$-functions
$\,$ (i) $\,$ $\be_1(a)$ $\,$ and 
$\,$ (ii) $\,$ $\be_3(a)$ for the case of the 
``super-simplified'' supersymmetry breaking model.}
\end{quotation}

\label{fig4}

\end{figure}

\end{center}

\begin{quotation}
\noindent
{\bf Figure 4.} The plots of the $\beta$-functions
$\,$ (i) $\,$ $\be_1(a)$ $\,$ and 
$\,$ (ii) $\,$ $\be_3(a)$ for the case of the 
``super-simplified'' supersymmetry breaking model.
\end{quotation}
\vskip 5mm

The $\,\be_3$-function for the parameter $\,a_3\,$ 
can be derived in the very same way as it was explained
in section 5. The final expression has the following form: 
\beq
\be_3^t\,=\,\frac{1}{(4\pi)^2}
\,\frac{(25a^2-68)}{18\,a^4}\,\cdot\,
\Big[\,3A\,(a^2-4)\,-\,a^2\,\Big]\,,
\label{SUSY-BoxR}
\eeq

The UV and IR limits of this expression are given by 
\beq
\be_3^{t\,UV}\,=\,-\,\frac{1}{(4\pi)^2}\,\cdot\,\frac49
\,+\,{\cal O}\Big(\frac{m^2}{p^2}\Big)\,,
\,\,\,\,\,\,\,\,\,\,\,\,\,\,\,\,\,\,
\be_3^{t\,IR}\,=\,\frac{1}{(4\pi)^2}\,\cdot\,\Big(\frac{17}{45}
\,-\,\frac{31\,a^2}{315}\Big)
\,+\,{\cal O}\Big(\frac{m^4}{p^4}\Big)\,\Big)\,.
\label{SUSY-BoxR IR and UV}
\eeq
It is easy to see that the sign of the $\,\be_3$-function
changes from negative in the UV to positive in the IR, as 
we should (of course) expect. The $\,\be_3$-function 
dependence on the momenta is smooth and
monotonic, as can be observed at the second plot at the 
Figure 4. We can see that the decoupling in the high 
derivative sectors goes in such a way that the 
transition between stable and unstable inflation 
\cite{insusy} performs in a smooth way. One can expect a
qualitatively similar behavior of the  $\,\be$-functions
in realistic models of supersymmetry. 

\section{Conclusions}

\quad\quad
We have performed the calculations of the effective action 
of vacuum for the massive scalar, fermion, and vector fields
to the second order in curvature, using a mass-dependent 
renormalization scheme. As a result, we
have found the explicit form of the decoupling of the 
massive fields in the higher derivative sector of the 
vacuum effective action. In the high-energy limit there
is a perfect correspondence between the $\be$-functions 
derived in a mass-dependent scheme and the standard ones
derived within the $\,{\overline {MS}}$-scheme. Also, in
the same limit we have established the correspondence with 
the anomaly-induced effective action derived for the 
massless conformal fields. 

In the low-energy limit the $\be$-functions for the 
massive fields tend to zero in a smooth monotone way.
For the supersymmetric model, the form of decoupling in 
the higher derivative sector indicates the possibility of 
the soft transition between the stable and unstable regimes 
in the anomaly-induced inflation \cite{insusy}. In particular, 
we can observe explicitly, using the ``super-simplified'' 
(but reliable) supersymmetry breaking model, how the 
decoupling of sparticles occurs. The detailed 
investigation of the cosmological applications 
will be reported separately. 
\vskip 8mm

\noindent
{\large\bf Acknowledgments.} 
The authors are grateful to N. Berkovits for stimulating 
conversation.
E.G. acknowledges warm hospitality of the Physics Department
at the Federal University of Juiz de Fora. The work of the 
authors has been supported by the research 
grant from FAPEMIG and (I.Sh.) by the fellowship from CNPq.

\newpage


\end{document}